# Study of Proposed Methods for Improving TCP Performance Over Wireless Links

Anshuman Sinha, *Schlumberger, Austin Technology Center*

TCP is designed for networks with assumption that major losses occur only due to congestion of network traffic. On a wireless network TCP misinterprets the transmission losses due to bit errors and handoffs as losses caused by congestion, and triggers congestion control mechanisms. Because of its end to end delivery model, congestion handling and avoidance mechanisms, TCP has been widely accepted as Transport layer protocol for internetworks. Extension of Internetworks over wireless links is inevitable with the spread of ubiquitous computing and mobile communications. This paper presents study of different mechanisms proposed to extend Transport Control Protocol and other alternate solutions to enhance end to end performance over lossy wireless links. The paper studies details of different design choices proposed and their technical advantages and disadvantages. Finally, an analysis and proposal for best choice of proposed schemes are made for wireless networks.

## 1 INTRODUCTION

WIRELESS internetworks with mobile computers like cellular phones, PDAs, tablet PCs and smartcards are inevitable in future. Wireless links have high bit error rate, high latencies and low bandwidth as compared to a wired link. The bit error is high because of its vulnerability to interference by gaussian and stray noise and signals. TCP on wireless triggers congestion avoidance mechanisms, wherein the congestion window is reduced exponentially, thereby reducing the effective window size. The exponential decline is to avoid further packet drops at the router which is causing the congestion. Wireless links have low bandwidth as compared to wired link, this is illustrated in the following table.

Wireless hosts usually move inside the wireless network which causes handoffs. For most cellular wireless networks, with an exception of CDMA, handoffs are short period of dis-connections, which occurs when a mobile host moves from one cell to another. These disconnections last between 10 to 100 milliseconds wherein the call is transferred either between MSC and BSCs. Often wireless networks are composed of two or more networks and mobile host switches between different networks which causes a longer split time than a cellular handoff typical. The mobile host moves frequently inside the network, which means recalculation of route to wireless host and handing over of the connection to new base station. The subsequent sections of paper are organized to present single most contributing factor in each section with analysis of proposed methods which use this technique.

TABLE 1

| Network Type | Bandwidth |
|---|---|
| Ethernet 10Base-T | 100 Mbps |
| Lucent Wavelan | 11 Mbps |
| Cisco Aironet | 11 Mbps |

March, 2003

## 2 SPLIT SECTION APPROACH

### 2.1 Indirect TCP

Indirect TCP (I-TCP) was the first split connection protocol proposed in [1]. Connection between fixed and mobile host is split into two, one between mobile host and base station and the other between base station and fixed host. TCP connection from fixed host to base station uses standard TCP, however the connection between base station and mobile host may be optimized for lossy wireless links. Separation of transport protocol stack at the base station splits congestion and flow control mechanisms between wired and wireless link. The end-to-end reliability is ensured by application layer using techniques elaborated further. Since split connection protocols do not guarantee reliable connection the application on wireless host as well as fixed host will have to be modified and relinked.

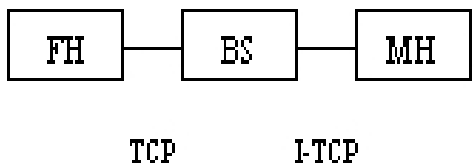

Fig. 1. I-TCP Protocol Scheme

On handoffs, connection state is passed from first base station to next base station to maintain connection between fixed host and mobile host. Base station has to maintain hard state of each connection between fixed and mobile host. This indirection is transparent from the fixed host. The handoff transfers I-TCP connection state between base stations, without affecting connection, since end points do not change in this technique.

I-TCP helps in disassociating wireless link connection with fixed host which shields fixed host from uncertainties of wireless link. In I-TCP, end to end semantics of TCP connection is lost since acknowledgements can arrive to sender evenbefore the actual packet has been delivered. During handoff, connection state at base station must be transferred to next base station. Usually, wireless link is the link with least bandwidth in complete connection from fixed to mobile host, hence base stations are complex and should have large buffer space in case of heavy traffic. If frequent handoffs occur, overhead related to connection state transfer between base stations may be large and might cause extra delays. Window from base station to wireless host is reduced on transmission errors. End-to-end latency overhead is added because the packet has to be copied from fixed host base station connection to base station mobile host buffer. Applications on wireless host have to be relinked with the I-TCP library to provide end-to-end reliable delivery.

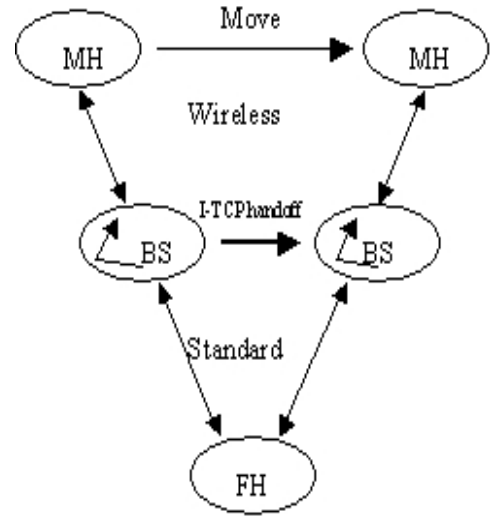

Fig. 2. I-TCP Connection Transfer Scheme

### 2.2 Selective Repeat Protocol

Selective Repeat Protocol (SRP) [2] is another example of split connection approach in which the connection is split at router adjacent to mobile host. Connection between fixed host and base station uses standard TCP, however transport layer protocol between base station and mobile host is modified for wireless links which selectively repeats. Unlike TCP, SRP has significant advantage of recovering more than one packet in single RTT. Selective acknowledgment (SACK) specifies missing segments using a bitmap. The base station upon receiving the SACK retransmits series of missing packet, which was lost in a burst. Another advantage of using SRP is realized when data is sent over from WH to FH. The MTU of a wireless link is usually much smaller than a wireline network. Using a normal TCP over wireless line would mean small MTU even for wired part of the route. On the other hand, using SRP, small segments are reassembled by MHP at the base station to take the advantage of larger MTU available over wired network. However, the base station has to maintain hard state and transfer the state to the next base-station when a handover occurs. The sender may receive acknowledge even before packet is delivered to the receiver.

# 3 LINK LAYER MECHANISMS

## 3.1 Forward Error Correction

Forward error correction (FEC) codes [16] are used to correct relatively small number of errors or erasures in byte streams. Erasures are easier to handle than error corrections since the exact position of loss is known. FEC hides correctable errors from TCP layer which reduces retransmissions from fixed host. The following figure shows the data encoding, decoding and reconstruction where n is the number of bit of original data and k is the size of error correction code.

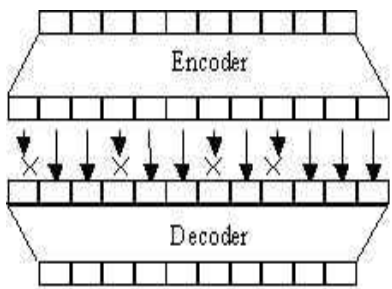

Fig. 3. Forward Error Correction

However, correction codes reduce effective bandwidth of wireless link and have additional load carried by packets in case of no error; partial throughput loss is compensated by eliminating need for sending dupacks. However, FEC needs additional computational power to verify the errors which is a constraint on mobile devices. Adaptive FEC reduces the overhead by choosing FEC codes dynamically depending on the error situation. FEC can be used in tandem with other link layer mechanisms to improve performance over wireless links. The link specific layers at base station and the mobile host needs to be changed to add FEC to each packet transmitted from base station to mobile.

## 3.2 Link Level Re-Transmissions

Link level retransmissions are used in conjunction with FEC, when the error rate is higher than correctable by FEC. The retransmissions are triggered by link level acknowledgements, duplicate acknowledgements, timeouts or negative acks. Retransmission takes less time as compared to the RTT of the connection. The retransmission at link level can cause congestion losses, since retransmission means reduction in bandwidth. Because of reduced bandwidth, there will be more packets in queue, due to which congestion control mechanisms may be triggered at fixed host. In case of few errors, RTT variations will be less and therefore, RTO will be smaller. The sender would retransmit the packet and so would the link layer retransmission scheme. Duplicate retransmissions waste the bandwidth and timeout at the sender, triggers congestion control mechanism. However, interference will not be an issue when the RTO granularity is high as compared to the RTT of the wireless link. Additionally, if the packets lost are too much out of order TCPs fast retransmit mechanisms at the sender will be triggered. This would again cause interference and waste of bandwidth due to duplicate retransmissions.

## 3.3 SNOOP Protocol

Snoop Protocol [4]: The snoop protocol introduces snoop agent at the base station, which snoops at each TCP packet going through the base station. Unacknowledged packets are cached at the BS for retransmission. If the receiver does not receive the packet or when the packet is out of order or on local timeout, the base station receives duplicate acknowledgement from the receiver. If the packet was cached at the base station, the base station retransmits the packets. The duplicate acknowledgment is absorbed by the snoop agent; therefore sender does not fast retransmit the out of order or lost packet. Advantage of snoop over other link layer mechanisms is the fact that it absorbs the duplicate acknowledgements. However, like other link layer mechanisms it suffers from possibility of interference and redundant duplicate retransmission. Snooping would fail in case when data and acknowledgment travel on different paths. End to end semantics of a TCP connection is maintained in the snoop protocol. The complete protocol stack need not process the packets since transmission is done at the link layer. During handoff the cached packets need to be transferred to the next base station. This is desirable but not necessary. The loss of cached packets will mean no loss recovery for short period just after however packets and acks have to travel the same path, that is go through the base station, so that base station can retransmit the packet. Snooping will not be useful if the packets are encrypted.

## 3.4 WTCP

WTCP [8]: Like snoop protocol, WTCP caches packets at the base station for retransmission, thereby hiding losses due to wireless link from the sender. In addition, WTCP uses the optional TCP extension of timestamp [RFC 1323] to estimate RTT. The base station adds base station residence time to the timestamp of the acknowledge. Adding the base station residence time to the timestamp neutralizes the affect of time spent by the packet at the base station. Adding base station timestamp shields senders RTT estimates due to wireless losses. Therefore, the RTO will not be affected and hence there wont be redundant duplicate retransmission by the sender. Timestamp option hides the time spent at the base station, such that the sender is not affected by the time spent in local retransmission due to wireless errors. TCP end to end semantics maintained. No modifications of network software at fixed or mobile host. Base station needs to maintain state and transfer at the time of handoff. The protocol assumes that the TCP extension of timestamp has been implemented at the sender, which is not widespread as yet.

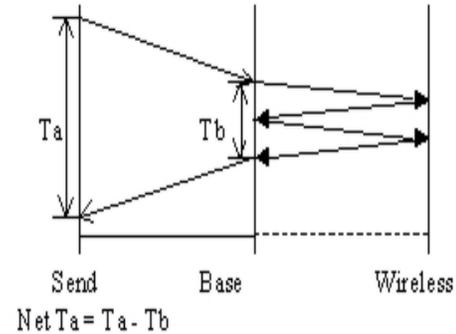

Fig. 4. Delayed Duplicate Acks

## 3.5 Delayed Duplicate Acks

Delayed duplicate acknowledge [15] tries to achieve two goals, one is to retransmit packet at the link layer and the other is stop the TCP [11] sender from transmitting duplicate packet which has already been supplied by the link layer segment retransmission at the base station. The receiver adds an extra delay before sending the third duplicate acknowledge, which is the last duplicate acknowledge before the sender does a fast retransmission. This gives link layer mechanism the extra time to retransmit the lost packet without causing the fast retransmit by the sender. This approach will be successful only in those cases when the wireless RTT is small when compared to the wired RTT, since the delay will most likely be insufficient to receive the retransmitted packet by the link layer protocol. The main advantage of this approach is that the link layer need not be TCP aware. TCP end to end semantics maintained and no modifications required at the fixed host. Recovery from congestion losses at link layer is delayed too, besides out-of-order recovery. Right value of delay depends on the link properties. Some kind of mechanism will be needed to guess the right delay between the second and the last duplicate acknowledge.

## 3.6    Fast Retransmission Approach

Fast Retransmit [14]: The fast retransmission approach can solve only the problem of handovers, which happens in most mobile communication networks, with exception of CDMA and its variants. The fixed host waits for the timeout before it can retransmit the packet. The handover consumes much less time than a typical roundtrip timeout; therefore it results in loss of end to end throughput. In the fast retransmission approach, the mobile host sends duplicate acknowledgements for packets lost due to handovers. The fixed host reduces the window and starts to retransmit without waiting for the timeout to occur. TCPs end to end semantics is maintained and no modifications are required at fixed host. Fast retransmission approach reduces handover delays and regains the lost throughput due to unnecessary timeouts. However, it deals only with handovers and not the error characteristics of the wireless link. This approach requires modification of the TCP code at the wireless host.

## 4    EXPLICIT NOTIFICATION MECHANISMS

### 4.1    ELN

ELN [7]: In this mechanism the base station records the losses in the packet sequence, which means it keeps track of the sequence number of lost packets. When the base station receives a duplicate acknowledge, it compares the duplicate acknowledge sequence number with the sequence number of packets that have passed this point in connection and sets the explicit loss notification (ELN) bit in the acknowledge. The sender upon receiving an acknowledge with ELN bit set, retransmits the packet without reducing the congestion window.

### 4.2    Explicit Bad State Notifications

EBSN [9]: The base station makes attempts to transmit the packet using link layer retransmission scheme. When it fails the base station sends an explicit bad state notification to the sender. The sender resets the timer, so that the packet is not retransmitted by the sender (FH) when the base station is still trying to retransmit the lost packet at the link layer.

## 4.3    Partial Acknowledge Protocols

Partial Acknowledge Protocols [12] [13]: The protocol implements two kinds of acknowledgment - partial and cumulative acknowledge. When the packet is received midway typically at base station, the base station sends a partial acknowledge ACKp to the sender. The sender upon receiving the partial acknowledgement ACKp does not trigger congestion control mechanisms. If the cumulative acknowledgement is not received by the sender, the packet is assumed to be lost due to transmission error on lossy wireless link. However, the congestion control mechanisms are triggered if either ACKp or ACKc is not received by the sender. Advantage of explicit notification is that fixed and the mobile host protocol stack need not be changed. The end-to-end semantics of TCP connection is maintained. Nonetheless connection state has to be maintained at base station and switched over to new station on handover.

## 4.4    End to End Mechanisms

End to end protocols have the following basic mechanisms. The receiver sends a selective acknowledgement upon loss of any packet on the wireless link. The sender interprets the selective acknowledgment and differentiates the loss from congestion with other losses, and implements Explicit Loss Notification mechanism. The reverse is also possible wherein the sender uses statistical calculations to predict the nature of loss.

## 4.5    Sender Based Discrimination

Sender based discrimination is based on statistical calculation from senders end. Besides sampling RTT for round trip timeout, the sender builds up a loss pattern. The loss pattern is determined by various statistical equations, which helps congestion window size. Different statistical algorithms have been used to deduce congestion window size. The disadvantage of this scheme is that the statistics collected by the sender is affected by traffic in the network which could be from an origin different than the sender, typically so in an internetwork. Also, the observance has to be done over long period to reliably predict the onset of congestion and hence calculate congestion window size. The current heuristics are primitive, and better ones needs to be developed. Probably the heuristics can be better guesses with advancement of the artificial intelligence science.

## 4.6 Receiver Based Discrimination

Mobile host, the receiver, assuming most of the data transfer occurs in this direction, analyzes the network traffic to predict the cause of transmission losses. The inter packet delay is analyzed to predict the reason for losses. On determining the reason using a heuristic the receiver sends either dupacks with ELN bit set, or explicit notification tag. The wireless link should be the last and slowest link in the network, so that some queuing takes place. The queuing delays at the receiver should be uniform, if otherwise the heuristic will not be able to deduce the right condition. However, with multiple connections at the base stations the queuing delays will be different for each connection causing the scheme to fail. This scheme is end-to-end; connection need not be split at the BS. The receiver-based discrimination maybe used even when the path taken by data and acks are not identical, and also when the data packets are encrypted.

## 5 CONCLUSION

The following table illustrates classification of the different proposed methods based on mechanisms used for increasing efficiency and the changes needed on the network stack at different locations in a connection. Different approaches have been taken to tackle the design problem, but the right technically correct changes is dependent on where the single or multiple modification needs to be done. This is particularly important for an existing system where the stack may need to be updated without upgrade of other system components. This paper gave the author opportunity to learn more about the different proposals. This was the primary objective for this paper. From the table below its clear that end to end protocols

TABLE 2

| | Sender Stack | Base Station | Receiver Stack |
|---|---|---|---|
| Split protocols: | | | |
| ITCP | | * | * |
| SRP | | * | * |
| Link layer mechanisms: | | | |
| Forward Error Correction | * | | * |
| Link layer retransmissions | * | * | |
| TCP aware mechanisms: | | | |
| SNOOP | | * | * |
| WTCP | | * | * |
| TCP unaware mechanisms: | | | |
| Delayed dupacks | | * | * |
| Fast Retransmit | * | | |
| Explicit Notifications: | | | |
| ELN | * | | * |
| EBSN | * | | * |
| Partial Ack Protocols | * | * | |
| End to End protocols: | | | |
| Sender discrimination | * | | |
| Receiver discrimination | | | * |

## ACKNOWLEDGMENTS

The author would like to thank everyone who studies TCP's extension for wireless links and proposed different methods. The author thanks all his colleagues who gave him the opportunity to learn and write about the topic. He would also like to thank his family for bearing with him while writing this technical paper. Special thanks to my father for inspiring me every moment of this life. A tribute to his great life and endeavor to make his children successful in their life.

# 6 APPENDIX

## TABLE 3

| Name | Form |
| --- | --- |
| BER | Bit Error Rate |
| ITCP | Indirect Transport Control Protocol |
| EBSN | Explicit State Bad Notification |
| SRP | Selective Repeat Protocol |
| SACK | Selective Acknowledgment |
| RTT | Round Trip Time |
| RTO | Retransmission Time out |
| MHP | Mobile Host Protocol |
| FH | Fixed Host |
| BS | Base Station |
| WH | Wireless Host |
| MH | Mobile Host |
| FEC | Forward Error Correction |
| ELN | Explicit Loss Notification |
| UDP | User Datagram Protocol |
| WTCP | Reliable TCP for Wide Area Wireless |

**Anshuman Sinha** Anshuman Sinha presently works for a global security company where he is responsible for smartcards and RFID based security and access control systems, which includes the hardware, firmware and software. He oversees the design, manufacturing and production of next generation contactless smartcards for his company. His primary expertise is in the area of near and far field secure RFID systems and Java based smartcards. He innovated the security and loading framework for Java Cards. He has several years of experience in developing Java card masks, on card software components, credit/debit, contactless, cryptography, Phase 2+ SIM and RFID applications. He has managed and led engineering teams for innovative product development and deployment. He has a passion to innovate and lead development of products that changes the industry and prevailing systems. Before his present role, he was software architect and team leader with global smart card and oilfield technology firm.